\documentclass{article}
\usepackage[preprint]{spconf,amsmath,graphicx}
\usepackage[hidelinks]{hyperref}

\title{Artificially Synthesising Data for Audio Classification and Segmentation to improve speech and music detection in Radio Broadcast}
\name{\begin{tabular}{c}Satvik Venkatesh$^{\star}$, David Moffat$^{\star}$, Alexis Kirke$^{\star}$, G\"{o}zel Shakeri$^{\dagger}$, Stephen Brewster$^{\dagger}$,\\J\"{o}rg Fachner$^{\ddagger}$, Helen Odell-Miller$^{\ddagger}$, Alex Street$^{\ddagger}$, Nicolas Farina$^{\dagger\dagger}$,\\Sube Banerjee$^{\ddagger\ddagger}$, and Eduardo Reck Miranda$^{\star}$\end{tabular}\thanks{This study was supported by Engineering and Physical Sciences Research Council (EPSRC) grants EP/S026991/1, EP/S027491/1, EP/S027203/1, and EP/S026959/1.}}

\address{\ninept$^{\star}$ Interdisciplinary Centre for Computer Music Research, University of Plymouth, UK\\
	\ninept$^{\dagger}$School of Computing Science, University of Glasgow, UK\\
	\ninept$^{\ddagger}$Cambridge Institute of Music Therapy Research, Anglia Ruskin University, UK\\
	\ninept$^{\dagger\dagger}$Centre for Dementia Studies, Brighton and Sussex Medical School, UK\\
	\ninept$^{\ddagger\ddagger}$Faculty of Health, University of Plymouth, UK}

\copyrightnotice{\begin{tabular}[t]{@{}l@{}}© 2021 IEEE. Personal use of this material is permitted. Permission from IEEE must be obtained for all other uses, in any current or\\future media, including reprinting/republishing this material for advertising or promotional purposes, creating new collective works,\\for resale or redistribution to servers or lists, or reuse of any copyrighted component of this work in other works.\end{tabular}}

\begin{document}
\ninept

\maketitle

\begin{abstract}
Segmenting audio into homogeneous sections such as music and speech helps us understand the content of audio. It is useful as a pre-processing step to index, store, and modify audio recordings, radio broadcasts and TV programmes. Deep learning models for segmentation are generally trained on copyrighted material, which cannot be shared. Annotating these datasets is time-consuming and expensive and therefore, it significantly slows down research progress. In this study, we present a novel procedure that artificially synthesises data that resembles radio signals. We replicate the workflow of a radio DJ in mixing audio and investigate parameters like fade curves and audio ducking. We trained a Convolutional Recurrent Neural Network (CRNN) on this synthesised data and outperformed state-of-the-art algorithms for music-speech detection. This paper demonstrates the data synthesis procedure as a highly effective technique to generate large datasets to train deep neural networks for audio segmentation.
\end{abstract}
\section{Introduction}
\label{sec:intro}

Automatically understanding the content of audio data is useful for indexing audio archives, target-based distribution of media, speech recognition, and intelligent remixing. It includes the task of audio segmentation, which divides an audio signal into homogeneous segments. These segments contain audio classes like music, speech, environmental sounds, and noise, to name but a few. The specificity of audio classes depends on the application. For instance, in radio broadcast, some relevant audio classes include music, speech, noise, and silence \cite{theodorou2014overview}.

Primarily, there are two approaches to audio segmentation --- (1) distance-based segmentation and (2) segmentation-by-classification \cite{butko2011audio}. In the former, boundaries of acoustic events are directly detected. This is done by calculating a distance metric, such as Euclidean distance, Bayesian information criterion (BIC) \cite{xue2010computationally}, or generalized likelihood ratio (GLR). For a given audio, a distance curve is plotted. The peaks on this distance curve are associated with the boundaries of audio events because they comprise high acoustic changes. The advantage of this technique is that it is generally unsupervised and does not require knowledge of the individual audio classes. However, the disadvantage is that it is more sensitive to dissimilarities within each audio class. 

In segmentation-by-classification, as the name suggests, the audio is divided into individual frames, typically in the range of 10 to 25 ms. These frames are independently classified and eventually the boundaries of audio events are detected. Traditionally, this was performed through algorithms like Gaussian mixture model (GMM) \cite{kos2009line}, support vector machine (SVM), and factor analysis (FA) \cite{castan2014audio}. In recent years, due to the advances in deep learning, segmentation-by-classification has gained more popularity through neural network architectures like bidirectional long short-term memory (B-LSTM) \cite{gimeno2020multiclass}, Convolutional Recurrent Neural Network (CRNN) \cite{choi2017convolutional}, and Temporal Convolutional Neural Network (TCN) \cite{lemaire2019temporal}.   

Machine learning models are generally trained using proprietary audio such as television and radio broadcast. This imposes a serious hindrance in the repeatability of research because this audio cannot be shared across different research groups. Annotating these datasets is a time-consuming and expensive task. For example, the study by Schl{\"u}ter et al. \cite{schluter2012unsupervised} annotated 42 hours of radio broadcast with the help of paid students. Moreover, a dataset called Open Broadcast Media Audio from TV (OpenBMAT) \cite{melendez2019open} was cross-annotated by three different annotators and each of them spent approximately 130 hours to annotate 27.4 hours of audio. As the labels in these datasets need to be precise enough for the models to train, the annotations in such datasets are generally verified by at least one other person. These factors impose many challenges for a researcher who wants to freshly explore audio segmentation.

The literature comprises many datasets that contain individual files of music and speech. However, these files are different from broadcast audio because broadcast audio is well-mixed. To our knowledge, the only openly available annotated database for this task is the MuSpeak dataset \cite{muspeak}, which contains approx. 5 hours of audio. Moreover, OpenBMAT \cite{melendez2019open} focused on estimating the relative loudness of music, but not speech and music detection.

In this paper, we present a novel approach to artificially synthesise audio that resembles a radio broadcast. We replicate the process of a radio DJ in mixing audio content. This was done by investigating fade curves, audio ducking, fade durations, and silences. The artificially mixed audio only uses openly available music-speech datasets that contain individual files of music and speech. Using this data synthesis procedure, large amounts of training data can be generated to train deep neural networks. The trained models are useful for real-world applications and achieve state-of-the-art performance on human-labelled datasets. The implementation, code, and pre-trained models associated with this study are openly available in this GitHub repository\footnote{\href{https://github.com/satvik-venkatesh/audio-seg-data-synth/}{https://github.com/satvik-venkatesh/audio-seg-data-synth/}}.


\section{Data Synthesis}
\label{sec:data-synthesis}

\subsection{Datasets}
In this study, we used datasets that contain audio files labelled as either music or speech. We did not use data that contains mixed audio. Instead, the radio content was synthesised through combining and mixing the music and speech data together. We used the MUSAN corpus \cite{snyder2015musan}, GTZAN music and speech detection dataset \cite{tzanetakis2000marsyas}, and the Scheirer \& Slaney dataset \cite{scheirer1997construction}. These are the commonly used datasets in music-speech detection studies \cite{lemaire2019temporal}. When we conducted initial tests with our neural network, we observed that there were confusions between wind instruments like flute and speech. Additionally, some vocal sections without accompaniment were confused with speech. Therefore, we extended our data repository to using the Instrument Recognition in Musical Audio Signals (IRMAS) dataset that includes many examples of wind instruments \cite{bosch2012comparison}, GTZAN genre recognition \cite{tzanetakis2002musical} for additional music examples, Singing Voice Audio Dataset which contains unaccompanied vocals \cite{black2014automatic}, and a section of the LibriSpeech corpus \cite{panayotov2015librispeech} for more speech examples. We also considered noise examples from the MUSAN corpus to enable the neural network to detect task-irrelevant examples. These are sounds that cannot be labelled as either music or speech. For instance, environmental sounds, babble noise, unintelligible speech, footsteps, and so on. The total number of audio files for music, speech, and noise was 6876, 6885, and 665 respectively. 

\subsection{Audio Transitions}
\begin{figure}[htb]
	
	\begin{minipage}[b]{0.48\linewidth}
		\centering
		\centerline{\includegraphics[width=4.0cm]{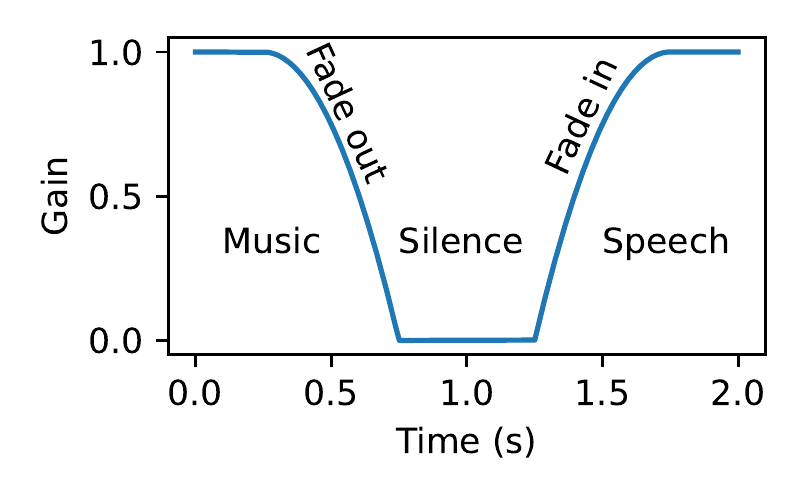}}
		\centerline{(a) Normal fade transition}\medskip
	\end{minipage}
	\begin{minipage}[b]{0.48\linewidth}
		\centering
		\centerline{\includegraphics[width=4.0cm]{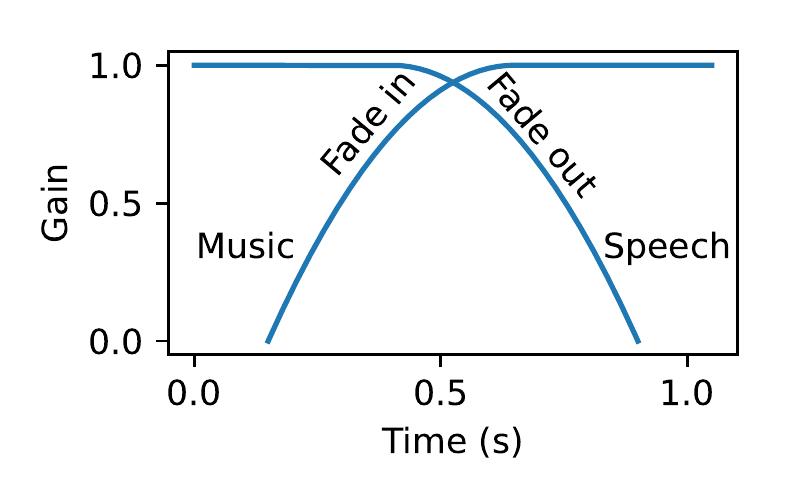}}
		\centerline{(b) Cross-fade transition}\medskip
	\end{minipage}
	\caption{Two types of audio transitions}
	\label{fig:fades}
\end{figure}

In radio programmes, shifts between music and speech and vice versa are generally smoothed through transitions. We broadly observed two types of transitions, which we have termed as normal transition and cross-fade transition. In a normal transition, an audio event is faded out, followed by a short period of silence, and then a new audio event is faded in. An example can be found in figure \ref{fig:fades}a. In a cross-fade transition, as the name suggests, the two audio signals are overlapping. While one is fading out, the other is fading in, as shown in figure \ref{fig:fades}b. 

\subsection{Fade Curves}
\begin{figure}[htb]
	\begin{minipage}[b]{1.0\linewidth}
		\centering
		\centerline{\includegraphics[width=8.5cm]{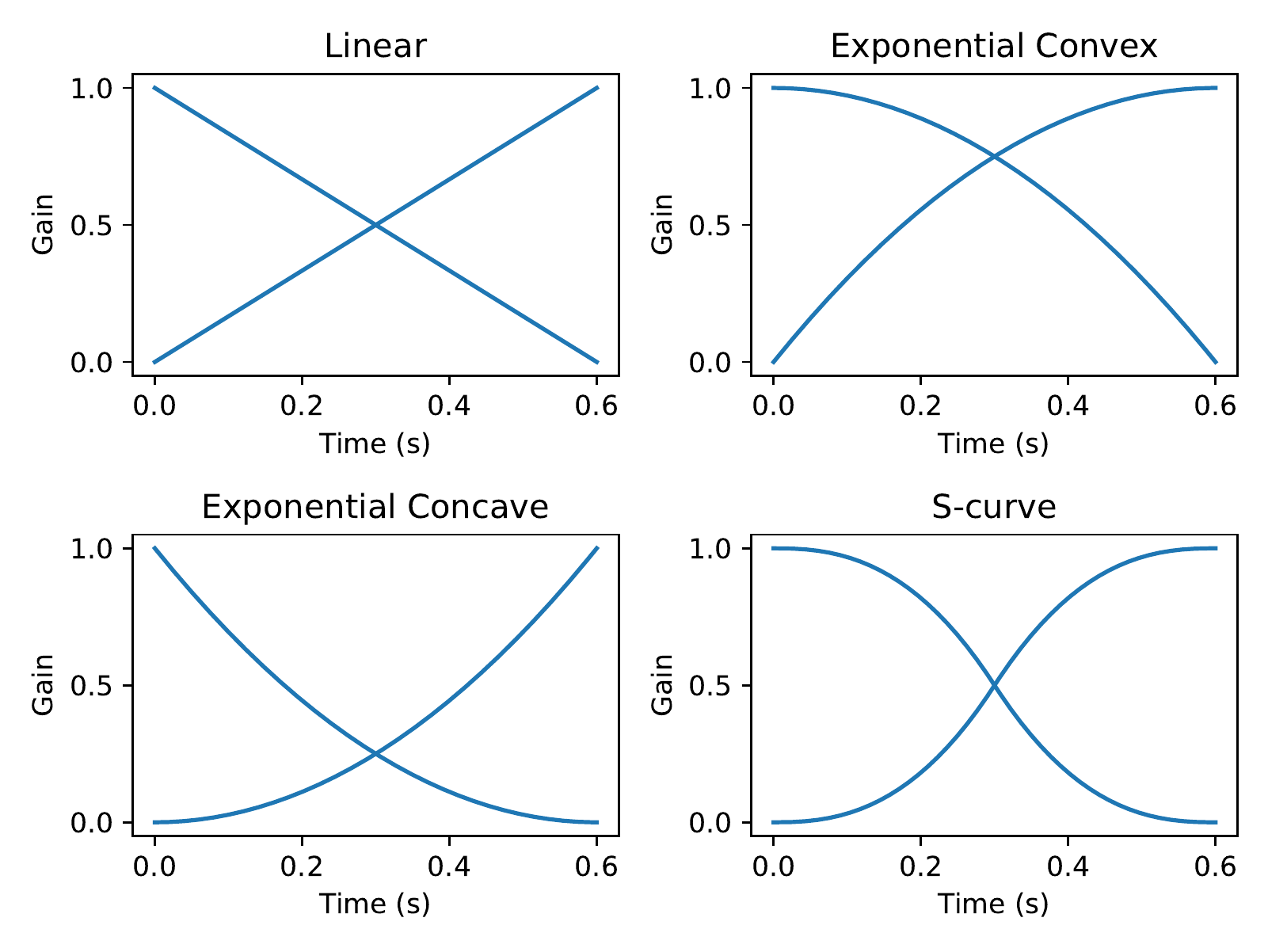}}
	\end{minipage}	
	\caption{Four types of fade curves}
	\label{fig:fade-curves}
\end{figure}

During audio mixing, engineers use different fade curves depending on the context. We have considered four popular curves that are commonly used in mixing \cite{tarr2018hack} --- linear, exponential convex, exponential concave, and s-curve. Figure \ref{fig:fade-curves} illustrates the types of fade curves.

\subsection{Parameters of Transitions}
Each audio example that was synthesised in this study was 8 s long. We felt that 8 s was a long enough duration to capture the entire fade curve. The three audio classes --- speech, music, and noise were stored in different directories. Each time an audio class was chosen for the data synthesis, a random file was selected from the entire list of files and then a random segment was extracted. 

For an audio transition, the two audio events and a time stamp of the transition are randomly chosen. For example, music to speech, speech to noise, speech to speech, etc., transitioning at a specific time. Note that we also allowed repetition of the same audio class, such as speech to speech because this would suggest cases like interviews. 

To cover a wide range of possibilities that can occur while mixing radio programmes, we randomised the various parameters of audio transitions. Each time an audio example was synthesised, a random fade curve was chosen. Subsequently, a random fade duration was chosen from a uniform distribution ranging from 0 s to the maximum possible duration. For a normal fade transition, the gap of silence between audio events was randomised.

\subsection{Background Music}
\label{sec:background-music}
In radio programmes, it is very common to have background music playing alongside foreground speech. Audio ducking is the process of reducing the volume of background music. It is generally performed to make speech intelligible. Many radio broadcasters have their own guidelines to audio ducking \cite{torcoli2019background}. Therefore, in order to artificially synthesise audio examples with background music, it is important for us to consider these guidelines.

We adopted the integrated loudness metric by ITU BS.1770-4 \cite{bs1770algorithms} to calculate the loudness of audio. This is measured in loudness units (LU). There is no ideal loudness difference (LD) between speech and background music because it is highly subjective. 
Commonly, the literature recommends a minimum LD of 7 to 10 LU \cite{torcoli2019background}. Moreover, in cases of very quiet background music, the LD can be as high as 23 LU. 

In order to implement our data synthesis procedure, we require a minimum and maximum LD to choose random values from a uniform distribution. 
We empirically observed the average performance of the network over multiple training cycles on different LDs and also manually listened to synthesised audio examples. We set the LD range to be between 7 and 18 LU.

In radio programmes, audio ducking can be performed through either volume automation or side-chain compression. We chose the former technique because it was relatively straightforward to achieve accurate LDs during data synthesis.

\subsection{Overview}
\begin{figure}[htb]
	\begin{minipage}[b]{1.0\linewidth}
		\centering
		\centerline{\includegraphics[width=8.5cm]{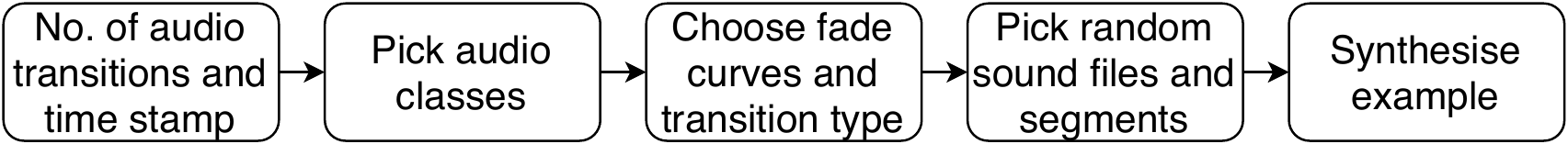}}
	\end{minipage}	
	\caption{An overview of the data synthesis procedure.}
	\label{fig:ds-flow}
\end{figure}

There are different combinations of audio classes that occur in the synthesised examples ---  music, speech, noise/silence, and speech over background music. Each example can either have no transitions (that is purely a single audio class) or one transition (that is two audio classes connected through fade curves) with a probability of 0.5. An overview of the data synthesis procedure can be found in figure \ref{fig:ds-flow}.

\section{Experiments}
\label{sec:experiments}

\subsection{Pre-processing and Feature Extraction}
All audio files used in this study were resampled to 22.05 kHz mono signals. Silences in the audio files were removed/shortened by using Sound eXchange (SoX). 
For our data synthesis to work smoothly, we required the audio files to have a minimum duration of 8 s. Some datasets such as the IRMAS have only 3 s audio files. To address this, we looped the audio to obtain the required duration.

Mel spectrograms have been commonly adopted by audio segmentation studies to extract features \cite{gimeno2020multiclass, lemaire2019temporal}. We set the hop size to 220 (10 ms) and FFT size to 1024 (46 ms). The selected audio segments were peak-normalised before synthesising an example. After synthesis, the whole example was peak-normalised again. We extracted 80 log-scale-Mel bands from 64 Hz to 8 kHz.

\subsection{Validation and Test Sets}
We are using artificially synthesised data for training. As these are not real-world examples, we cannot use synthesised data for validation and testing. We incorporated the MuSpeak dataset, which contains 5 h 14 m 14 s of audio. We also collected 9 h of broadcast audio from BBC Radio Devon, which was manually annotated by the authors. The audio in the BBC dataset were split into files of 1 h. 
Three hours of our annotations were verified by an external audio mixing engineer who was not involved in the research and paid for his time. Additionally, a random section of 15 minutes was blind-annotated by him independently. We found an agreement of 99.49\% with our annotations by using 10 ms segment verifications. This was done to ensure that audio events were similarly perceived by different people. 
In order to explore the robustness of data synthesis, we did not use any of this data for training. The data from MuSpeak and BBC dataset was shuffled and used as validation and test sets as a 50-50\% split.

In order to compare our model with other state-of-the-art algorithms, we also evaluated it on dataset number 1 of the Music Information Retrieval Evaluation eXchange (MIREX) 2018 music and speech detection competition\footnote{\href{https://music-ir.org/mirex/wiki/2018:Music\_and/or\_Speech\_Detection}{https://music-ir.org/mirex/wiki/2018:Music\_and/or\_Speech\_Detection}}. This dataset contains 27 hours of audio from various TV programmes. 
Although our data synthesis was designed for radio programmes, this dataset would provide us with a good evaluation of our model.

\subsection{Network Architecture}

For this study, we adopted a CRNN, which is a state-of-the-art architecture for audio classification and segmentation tasks \cite{choi2017convolutional, cakir2017convolutional}. The input shape of the network was $802\times80\times1$, equivalent to 802 time steps and 80 Mel bins. The output of the network comprised $802\times2$ neurons with sigmoid activations, where two neurons perform a binary classification for music and speech at every time step. The network performs multi-output detection, independently detecting the regions of music and speech. This is important for models working with radio data because music and speech can occur simultaneously. Binary cross-entropy was used as the loss function.

We used the Adam optimizer with a constant learning rate of 0.001 and batch size of 128. The first two layers of the network were 2D convolutional layers with a kernel size of 7 and a stride of 1. The input was padded  with zeros such that `same' convolutions were performed to ensure that the time resolution remains the same. The next two layers were bidirectional gated recurrent units (B-GRU) with 80 units each. 

In this study, we evaluated the model using different training sets, as explained in section \ref{sec:training-datasets}. Hence, a model architecture was finalised by optimising the performance across different training datasets. For regularisation, we implemented early-stopping and used batch normalisation after all the layers. Max pooling along the dimension of Mel bins was performed after the convolutional layers. A dropout of 0.2 was added only after the convolutional layers because we observed that it was not effective for the B-GRU layers.

\subsection{Training Datasets}
\label{sec:training-datasets}
In order to evaluate the effectiveness of our data synthesis algorithm, we constructed 4 training datasets. All datasets contain 40960 examples of 8 s audio (which is approximately 91 h of audio). Initial tests conveyed this was an adequate number of examples to train the network. 
\begin{enumerate}
	\item Dataset-only files (d-OF): This dataset contains audio segments of only speech, music, or noise. There was no mixing of audio events within each example. 40960 examples were randomly sampled from our data repository. We did not include the whole corpus because of computational limitations and to manage redundancy.
	
	\item Dataset-only files and background music (d-OFB): In addition to d-OF, this dataset contains examples of speech over background music. The volume of background music was normalised according to the method explained in section \ref{sec:background-music}. However, this dataset did not contain any audio transitions.
	
	\item Dataset-no normalisation (d-NN): In this dataset, the data synthesis was performed as explained in section \ref{sec:data-synthesis}, except for the loudness normalisation of background music according to loudness of foreground speech. However, all examples of speech, music, and noise were peak-normalised before synthesis. 
 
	\item Dataset-data synthesis (d-DS): In this dataset, the data synthesis was performed exactly as explained in section \ref{sec:data-synthesis}.
	\end{enumerate}

\subsection{Post-processing}
A threshold of 0.5 was used to make binary classifications on the output layer. The length of each file in the test set was approximately 1 h. We traversed the audio file with a window size of 8 s and hop size of 6 s. We discarded the predictions made on the first and last second of each audio example because they might be unreliable. This technique was adopted from the study by Gimeno et al. \cite{gimeno2020multiclass}.

In the audio segmentation pipeline, predictions made by the model are generally sent though a post-processing phase to remove spurious transitions and events. This is done through either median filtering \cite{gimeno2020multiclass, schluter2012unsupervised} or setting thresholds for minimum durations of audio events \cite{lemaire2019temporal}. We adopted the latter approach and set thresholds for minimum speech duration, minimum music duration, maximum silence between speech, and maximum silence between music. These values were obtained from the study by Lemaire et al. \cite{lemaire2019temporal} and set to 1.3 s, 3.4 s, 0.4 s, and 0.6 s respectively.  

\section{Results}
To evaluate the models, we adopted metrics implemented in the sed\_eval toolbox \cite{mesaros2016metrics}, which has been widely adopted by audio event detection studies \cite{lemaire2019temporal, cakir2017convolutional, mesaros2017dcase}. The segment-level evaluation was performed with a segment size of 10 ms. Table \ref{table:eval} presents the model's performance on different datasets. The highest overall F-measure was obtained by d-DS, which implemented the entire data synthesis procedure. The F-measures of d-OF and d-OFB were at least 3\% lower than d-DS because their datasets did not contain audio transitions. This demonstrates that modelling radio DJ-like transitions is an effective technique. Additionally, there is a marginal difference between d-OF and d-OFB, which explains that adding background music to speech in the training examples is not sufficient, but there needs to be audio transitions.

The dataset d-NN contained background music that was peak-normalised, but not normalised with respect to loudness of foreground speech. Therefore, music F-measure of d-DS surpasses the value of d-NN by more than 2\%. This proves that randomising the loudness of background music with respect to foreground speech within a LD of 7 to 18 was an effective method. Speech F-measure for d-NN was slightly greater than d-DS. However, this might be because the background music in d-NN was at a relatively constant volume, which improves speech detection but compromises music detection. 

\begin{table}[htb]
	\centering
	\begin{tabular}{|c|c|c|c|} \hline
		\textbf{Dataset} & \textbf{F\textsubscript{overall}} & \textbf{F\textsubscript{s}} & \textbf{F\textsubscript{m}} \\
		\hline
		d-OF    & 93.54 & 94.58 & 92.99 \\
		d-OFB   & 93.68 & 94.95 & 92.99 \\ 
		d-NN & 95.33 & \textbf{96.44} & 94.73 \\
		d-DS    & \textbf{96.69} & 96.17 & \textbf{96.97} \\
		\hline
	\end{tabular}
\caption{The F-measure of our CRNN model trained on different datasets.}
\label{table:eval}
\end{table}

Table \ref{table:mirex} shows the segment-level evaluation of our d-DS model on the MIREX speech and music detection dataset. The evaluations of other submissions were obtained from the MIREX website. Our model significantly outperforms the other models for F-measures of music. This is attributed to the presence of audio transitions and loudness normalisation of background music in the synthesised dataset. Our model also obtains the highest F-measure for speech detection. 

All the other submissions in the competition used real-world data \cite{choi2018hybrid,marolt2018music}. Therefore, these results demonstrate that our data synthesis is a highly effective approach for audio segmentation. Moreover, there was another task in MIREX 2018 that was solely for music detection. Our model places second in this task, preceded by the submission by Mel{\'e}ndez-Catal{\'a}n et al. \cite{melendez2018music}. Their model was trained on 30 hours of TV programmes, which comes from the same data distribution. It is important to note that the MIREX evaluation dataset can contain background music over foreground speech, audience noises, sound effects, everyday-life sounds, sounds of the city, and so on. As our data synthesis procedure only considered foreground speech, it explains the poor precision for music in table \ref{table:mirex}. Our model predicted many of the sound effects as music. The performance of our model over TV programmes can be improved by considering these factors in the data synthesis.

\begin{table}[htb]
	\centering
	\begin{tabular}{|c|c|c|c|c|c|c|} \hline
		\textbf{Algo.} & \textbf{F\textsubscript{m}} & \textbf{P\textsubscript{m}} & \textbf{R\textsubscript{m}} & \textbf{F\textsubscript{s}} & \textbf{P\textsubscript{s}} & \textbf{R\textsubscript{s}} \\
		\hline
		\cite{choi2018hybrid}     & 49.36          & 62.4           & 40.82          & 77.18          & \textbf{96.83} & 64.15          \\
		\cite[a]{marolt2018music} & 38.99          & 80.72          & 25.7           & 91.15          & 87.95          & 94.6           \\
		\cite[b]{marolt2018music} & 54.78          & 85.7           & 40.26          & 90.9           & 89.45          & 92.41          \\
		\cite[c]{marolt2018music} & 31.24          & \textbf{98.73} & 18.56          & 90.86          & 83.83          & \textbf{99.17} \\
		d-DS                      & \textbf{85.76} & 79.37          & \textbf{93.27} & \textbf{92.21} & 89.71          & 94.85     		 \\ \hline
	\end{tabular}
	\caption{F-measure, precision, and recall of our CRNN model trained on `d-DS' and other algorithms evaluated on dataset number 1 of MIREX 2018 speech and music detection competition.}
	\label{table:mirex}
\end{table}

\section{Concluding Discussions}
In this study, only artificially synthesised data was used to train a model for audio segmentation and classification. We adopted a training dataset belonging to a different distribution from the validation and test sets. Despite this, we obtained a high F-measure on our local test set. Furthermore, we obtained state-of-the-art performance for speech and music detection on the MIREX 2018 competition dataset.

There were noticeable differences between the BBC Radio Devon recordings and the data repository we have used for data synthesis. The BBC recordings have greater dynamic range compression, cleaner speech, and generally use side-chain compression for audio ducking. Therefore, including a small number of radio recordings in the training dataset might improve the model's performance. Additionally, incorporating audio effects like dynamic range compression in the data synthesis pipeline might improve the model's performance.

Many studies have suggested end-to-end deep learning to be a potential pathway for future audio classification and segmentation research \cite{lemaire2019temporal, lee2017raw}. However, it requires much more data than using Mel spectrograms as features. As labelling large amounts of data is an expensive and time-consuming task, our data synthesis procedure serves as a potential solution to generate large amounts of training data and advance the state-of-the-art in audio segmentation and classification systems.

\vfill\pagebreak

\bibliographystyle{IEEEbib}
\bibliography{data-synth-audio-seg}

\begin{thebibliography}{10}

\bibitem{theodorou2014overview}
Theodoros Theodorou, Iosif Mporas, and Nikos Fakotakis,
\newblock ``An overview of automatic audio segmentation,''
\newblock {\em International Journal of Information Technology and Computer
  Science (IJITCS)}, vol. 6, no. 11, pp. 1, 2014.

\bibitem{butko2011audio}
Taras Butko and Climent Nadeu,
\newblock ``Audio segmentation of broadcast news in the albayzin-2010
  evaluation: overview, results, and discussion,''
\newblock {\em EURASIP Journal on Audio, Speech, and Music Processing}, vol.
  2011, no. 1, pp. 1, 2011.

\bibitem{xue2010computationally}
Hao Xue, HaiFeng Li, Chang Gao, and ZiQiang Shi,
\newblock ``Computationally efficient audio segmentation through a multi-stage
  {BIC} approach,''
\newblock in {\em 3rd International Congress on Image and Signal Processing}.
  IEEE, 2010, vol.~8, pp. 3774--3777.

\bibitem{kos2009line}
Marko Kos, Matej Grasic, Damjan Vlaj, and Zdravko Kacic,
\newblock ``On-line speech/music segmentation for broadcast news domain,''
\newblock in {\em 2009 16th International Conference on Systems, Signals and
  Image Processing}. IEEE, 2009, pp. 1--4.

\bibitem{castan2014audio}
Diego Cast{\'a}n, Alfonso Ortega, Antonio Miguel, and Eduardo Lleida,
\newblock ``Audio segmentation-by-classification approach based on factor
  analysis in broadcast news domain,''
\newblock {\em EURASIP Journal on Audio, Speech, and Music Processing}, vol.
  2014, no. 1, pp. 34, 2014.

\bibitem{gimeno2020multiclass}
Pablo Gimeno, Ignacio Vi{\~n}als, Alfonso Ortega, Antonio Miguel, and Eduardo
  Lleida,
\newblock ``Multiclass audio segmentation based on recurrent neural networks
  for broadcast domain data,''
\newblock {\em EURASIP Journal on Audio, Speech, and Music Processing}, vol.
  2020, no. 1, pp. 1--19, 2020.

\bibitem{choi2017convolutional}
Keunwoo Choi, Gy{\"o}rgy Fazekas, Mark Sandler, and Kyunghyun Cho,
\newblock ``Convolutional recurrent neural networks for music classification,''
\newblock in {\em IEEE International Conference on Acoustics, Speech and Signal
  Processing (ICASSP)}, 2017, pp. 2392--2396.

\bibitem{lemaire2019temporal}
Quentin Lemaire and Andre Holzapfel,
\newblock ``Temporal convolutional networks for speech and music detection in
  radio broadcast,''
\newblock in {\em 20th International Society for Music Information Retrieval
  Conference (ISMIR)}, 2019.

\bibitem{schluter2012unsupervised}
Jan Schl{\"u}ter and Reinhard Sonnleitner,
\newblock ``Unsupervised feature learning for speech and music detection in
  radio broadcasts,''
\newblock in {\em Proceedings of the 15th International Conference on Digital
  Audio Effects (DAFx)}, 2012.

\bibitem{melendez2019open}
Blai Mel{\'e}ndez-Catal{\'a}n, Emilio Molina, and Emilia G{\'o}mez,
\newblock ``Open broadcast media audio from {TV}: A dataset of tv broadcast
  audio with relative music loudness annotations,''
\newblock {\em Transactions of the International Society for Music Information
  Retrieval}, vol. 2, no. 1, 2019.

\bibitem{muspeak}
{MuSpeak Team},
\newblock ``{MIREX} muspeak sample dataset,'' 2015,
\newblock \url{http://mirg.city.ac.uk/datasets/muspeak/} [Last accessed on
  13-10-2020].

\bibitem{snyder2015musan}
David Snyder, Guoguo Chen, and Daniel Povey,
\newblock ``Musan: A music, speech, and noise corpus,''
\newblock {\em arXiv preprint arXiv:1510.08484}, 2015.

\bibitem{tzanetakis2000marsyas}
George Tzanetakis and Perry Cook,
\newblock ``Marsyas: A framework for audio analysis,''
\newblock {\em Organised sound}, vol. 4, no. 3, pp. 169--175, 2000.

\bibitem{scheirer1997construction}
Eric Scheirer and Malcolm Slaney,
\newblock ``Construction and evaluation of a robust multifeature speech/music
  discriminator,''
\newblock in {\em IEEE International Conference on Acoustics, Speech, and
  Signal processing (ICASSP)}, 1997, vol.~2, pp. 1331--1334.

\bibitem{bosch2012comparison}
Juan~J Bosch, Jordi Janer, Ferdinand Fuhrmann, and Perfecto Herrera,
\newblock ``A comparison of sound segregation techniques for predominant
  instrument recognition in musical audio signals.,''
\newblock in {\em 13th International Society for Music Information Retrieval
  Conference (ISMIR)}, 2012, pp. 559--564.

\bibitem{tzanetakis2002musical}
George Tzanetakis and Perry Cook,
\newblock ``Musical genre classification of audio signals,''
\newblock {\em IEEE Transactions on Speech and Audio processing}, vol. 10, no.
  5, pp. 293--302, 2002.

\bibitem{black2014automatic}
Dawn A.~A. Black, Ma~Li, and Mi~Tian,
\newblock ``Automatic identification of emotional cues in chinese opera
  singing,''
\newblock in {\em 13th International Conference on Music Perception and
  Cognition and the 5th Conference for the Asian-Pacific Society for Cognitive
  Sciences of Music}, 2014.

\bibitem{panayotov2015librispeech}
Vassil Panayotov, Guoguo Chen, Daniel Povey, and Sanjeev Khudanpur,
\newblock ``{L}ibri{S}peech: an {ASR} corpus based on public domain audio
  books,''
\newblock in {\em IEEE International Conference on Acoustics, Speech and Signal
  Processing (ICASSP)}, 2015, pp. 5206--5210.

\bibitem{tarr2018hack}
Eric Tarr,
\newblock {\em Hack Audio: An Introduction to Computer Programming and Digital
  Signal Processing in MATLAB},
\newblock Routledge, 2018.

\bibitem{torcoli2019background}
Matteo Torcoli, Alex Freke-Morin, Jouni Paulus, Christian Simon, and Ben
  Shirley,
\newblock ``Background ducking to produce esthetically pleasing audio for {TV}
  with clear speech,''
\newblock in {\em 146th Audio Engineering Society Convention}, 2019.

\bibitem{bs1770algorithms}
{ITU-R},
\newblock ``{ITU-R Rec. BS.1770-4}: Algorithms to measure audio programme
  loudness and true-peak audio level,''
\newblock 2017.

\bibitem{cakir2017convolutional}
Emre Cak{\i}r, Giambattista Parascandolo, Toni Heittola, Heikki Huttunen, and
  Tuomas Virtanen,
\newblock ``Convolutional recurrent neural networks for polyphonic sound event
  detection,''
\newblock {\em IEEE/ACM Transactions on Audio, Speech, and Language
  Processing}, vol. 25, no. 6, pp. 1291--1303, 2017.

\bibitem{mesaros2016metrics}
Annamaria Mesaros, Toni Heittola, and Tuomas Virtanen,
\newblock ``Metrics for polyphonic sound event detection,''
\newblock {\em Applied Sciences}, vol. 6, no. 6, pp. 162, 2016.

\bibitem{mesaros2017dcase}
Annamaria Mesaros, Toni Heittola, Aleksandr Diment, Benjamin Elizalde, Ankit
  Shah, Emmanuel Vincent, Bhiksha Raj, and Tuomas Virtanen,
\newblock ``{DCASE} 2017 challenge setup: Tasks, datasets and baseline
  system,''
\newblock in {\em Workshop on Detection and Classification of Acoustic Scenes
  and Events}, 2017.

\bibitem{choi2018hybrid}
Minsuk Choi, Jongpil Lee, and Juhan Nam,
\newblock ``Hybrid features for music and speech detection,''
\newblock {\em Music Information Retrieval Evaluation eXchange (MIREX)}, 2018.

\bibitem{marolt2018music}
Matija Marolt,
\newblock ``Music/speech classification and detection submission for {MIREX}
  2018,''
\newblock {\em Music Information Retrieval Evaluation eXchange (MIREX)}, 2018.

\bibitem{melendez2018music}
Blai Mel{\'e}ndez-Catal{\'a}n, E~Molina, and E~Gomez,
\newblock ``Music and/or speech detection {MIREX} 2018 submission,''
\newblock {\em Music Information Retrieval Evaluation eXchange (MIREX)}, 2018.

\bibitem{lee2017raw}
Jongpil Lee, Taejun Kim, Jiyoung Park, and Juhan Nam,
\newblock ``Raw waveform-based audio classification using sample-level {CNN}
  architectures,''
\newblock {\em 31st Conference on Neural Information Processing Systems
  (NIPS)}, 2017.

\end{thebibliography}

\end{document}